\documentstyle[12pt,epsfig]{article}

\newcommand{\agt}{\,\rlap{\lower 3.5 pt \hbox{$\mathchar \sim$}} \raise 1pt
 \hbox {$>$}\,}
\newcommand{\alt}{\,\rlap{\lower 3.5 pt \hbox{$\mathchar \sim$}} \raise 1pt
 \hbox {$<$}\,}

\catcode`@=11
\newcount\@tempcntc
\def\@citex[#1]#2{\if@filesw\immediate\write\@auxout{\string\citation{#2}}\fi
  \@tempcnta\z@\@tempcntb\m@ne\def\@citea{}\@cite{\@for\@citeb:=#2\do
    {\@ifundefined
       {b@\@citeb}{\@citeo\@tempcntb\m@ne\@citea\def\@citea{,}{\bf ?}\@warning
       {Citation `\@citeb' on page \thepage \space undefined}}%
    {\setbox\z@\hbox{\global\@tempcntc0\csname b@\@citeb\endcsname\relax}%
     \ifnum\@tempcntc=\z@ \@citeo\@tempcntb\m@ne
       \@citea\def\@citea{,}\hbox{\csname b@\@citeb\endcsname}%
     \else
      \advance\@tempcntb\@ne
      \ifnum\@tempcntb=\@tempcntc
      \else\advance\@tempcntb\m@ne\@citeo
      \@tempcnta\@tempcntc\@tempcntb\@tempcntc\fi\fi}}\@citeo}{#1}}
\def\@citeo{\ifnum\@tempcnta>\@tempcntb\else\@citea\def\@citea{,}%
  \ifnum\@tempcnta=\@tempcntb\the\@tempcnta\else
   {\advance\@tempcnta\@ne\ifnum\@tempcnta=\@tempcntb \else \def\@citea{--}\fi
    \advance\@tempcnta\m@ne\the\@tempcnta\@citea\the\@tempcntb}\fi\fi}
\catcode`@=12

\begin{document}

\title{\vskip-3cm{\baselineskip14pt
\centerline{\normalsize DESY 00-110\hfill ISSN 0418-9833}
\centerline{\normalsize MPI-PhT/2000-27\hfill}
\centerline{\normalsize hep-ph/0007336\hfill}
\centerline{\normalsize July 2000\hfill}
}
\vskip1.5cm
Squark Loop Correction to $W^\pm H^\mp$ Associated Hadroproduction
}
\author{A. A. Barrientos Bendez\'u$^1$ and B. A. Kniehl$^{2,}$\thanks{%
Permanent address: II. Institut f\"ur Theoretische Physik, Universit\"at
Hamburg, Luruper Chaussee 149, 22761 Hamburg, Germany.}\\
{\normalsize $^1$
II. Institut f\"ur Theoretische Physik, Universit\"at Hamburg,}\\
{\normalsize Luruper Chaussee 149, 22761 Hamburg, Germany}\\
{\normalsize $^2$
Max-Planck-Institut f\"ur Physik (Werner-Heisenberg-Institut),}\\
{\normalsize F\"ohringer Ring 6, 80805 Munich, Germany}}

\date{}

\maketitle

\thispagestyle{empty}

\begin{abstract}
We study the squark loop correction to $W^\pm H^\mp$ associated
hadroproduction via gluon-gluon fusion within the minimal supersymmetric
extension of the standard model.
We list full analytic results and quantitatively analyze the resulting shift
in the cross section at the CERN Large Hadron Collider assuming a
supergravity-inspired scenario.

\medskip

\end{abstract}

\newpage

\section{Introduction}

The search for Higgs bosons will be among the prime tasks of the CERN Large
Hadron Collider (LHC) \cite{kun}.
While the standard model (SM) contains one complex Higgs doublet, from which
one neutral CP-even Higgs boson emerges in the physical particle spectrum
after the electroweak symmetry breaking, the Higgs sector of the minimal
supersymmetric extension of the SM (MSSM) consists of a two-Higgs-doublet
model (2HDM) and accommodates five physical Higgs bosons:
the neutral CP-even $h^0$ and $H^0$ bosons, the neutral CP-odd $A^0$ boson,
and the charged $H^\pm$-boson pair.
At the tree level, the MSSM Higgs sector has two free parameters, which are
usually taken to be the mass $m_A$ of the $A^0$ boson and the ratio
$\tan\beta=v_2/v_1$ of the vacuum expectation values of the two Higgs
doublets.

The discovery of the $H^\pm$ bosons would rule out the SM and, at the same 
time, give strong support to the MSSM.
The main strategies for the $H^\pm$-boson search at the LHC were summarized in
Refs.~\cite{kun,wh}.
Depending on the $H^\pm$-boson mass $m_H$, the dominant mechanism of single
$H^\pm$-boson hadroproduction are
$gg,q\bar q\to t\bar t$ followed by $t\to bH^+$ \cite{kun},
$g\bar b\to\bar tH^+$ \cite{gun}, $gg\to\bar tbH^+$ \cite{dia},
and $qb\to q^\prime bH^+$ \cite{mor} together with their charge-conjugate
counterparts.
The hadroproduction of $H^+H^-$ pairs proceeds at tree the level via $q\bar q$
annihilation, $q\bar q\to H^+H^-$, where $q=u,d,s,c$ \cite{eic}, and $b$ 
\cite{hh}, and at the one-loop level via $gg$ fusion, $gg\to H^+H^-$, which is
mediated by quark \cite{hh,wil,bre} and squark loops \cite{hh,bre}.
The suppression of the $gg$-fusion cross section by two powers of the
strong-coupling constant $\alpha_s$ relative to the one of $q\bar q$
annihilation is partly compensated at multi-TeV hadron colliders by the
overwhelming gluon luminosity.

\begin{figure}[ht]
\begin{center}
\centerline{\epsfig{figure=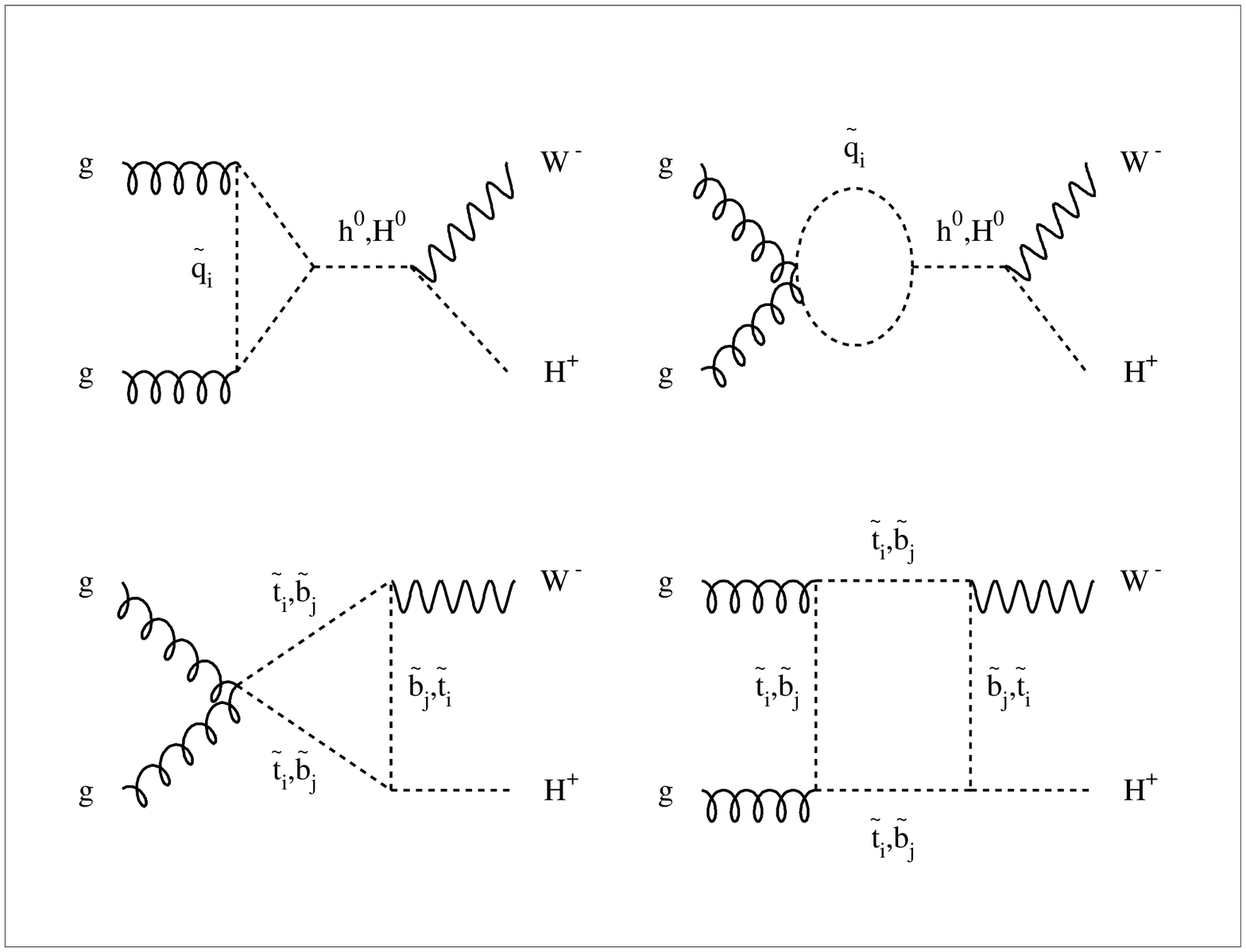,width=16cm}}
\caption{One-loop Feynman diagrams for $gg\to W^-H^+$ due to virtual squarks
in the MSSM.}
\label{fig:one}
\end{center}
\end{figure}

An interesting alternative is to produce $H^\pm$ bosons in association with
$W^\mp$ bosons, so that the leptonic decays of the latter may serve as a
trigger for the $H^\pm$-boson search.
The dominant partonic subprocesses of $W^\pm H^\mp$ associated production are
$b\bar b\to W^\pm H^\mp$ at the tree level and $gg\to W^\pm H^\mp$ at one 
loop, which were investigated for vanishing bottom-quark mass $m_b$ and small
values of $\tan\beta$ ($0.3\le\tan\beta\le2.3$) in Ref.~\cite{dic} and
recently, without these restrictions, in Refs.~\cite{wh,box}.
A careful signal-versus-background analysis, based on the analytic results of
Ref.~\cite{wh}, was recently reported in Ref.~\cite{smo}.
So far, only the quark loop contribution to $gg\to W^\pm H^\mp$ was
considered \cite{wh,dic,box}.
The purpose of this paper is to provide, in analytic form, the supersymmetric
contribution to this partonic subprocess, which is induced by virtual squarks
through the Feynman diagrams depicted in Fig.~\ref{fig:one}.
Furthermore, we wish to quantitatively study its influence on the cross
section of the inclusive reaction $pp\to W^\pm H^\mp+X$ at the LHC.
We recall that, in the case of $pp\to H^+H^-+X$, the supersymmetric correction
to the $gg$-fusion cross section can be as large as $+50\%$ \cite{hh}.
{\it A priori}, one expects to encounter a similar situation for
$pp\to W^\pm H^\mp+X$.

In order to reduce the number of unknown supersymmetric input parameters, we 
adopt a scenario where the MSSM is embedded in a grand unified theory (GUT)
involving supergravity (SUGRA) \cite{kal}.
The MSSM thus constrained is characterized by the following parameters at the
GUT scale, which come in addition to $\tan\beta$ and $m_A$: the universal
scalar mass $m_0$, the universal gaugino mass $m_{1/2}$, the trilinear
Higgs-sfermion coupling $A$, the bilinear Higgs coupling $B$, and the
Higgs-higgsino mass parameter $\mu$.
Notice that $m_A$ is then not an independent parameter anymore, but it is
fixed through the renormalization group equation.
The number of parameters can be further reduced by making additional
assumptions.
Unification of the $\tau$-lepton and $b$-quark Yukawa couplings at the GUT
scale leads to a correlation between $m_t$ and $\tan\beta$.
Furthermore, if the electroweak symmetry is broken radiatively, then $B$ and
$\mu$ are determined up to the sign of $\mu$.
Finally, it turns out that the MSSM parameters are nearly independent of the
value of $A$, as long as $|A|\alt500$~GeV at the GUT scale.

This paper is organized as follows.
In Sec.~\ref{sec:two}, we list the helicity amplitudes of the partonic
subprocess $gg\to W^-H^+$ involving virtual squarks.
In Sec.~\ref{sec:three}, we present quantitative predictions for the inclusive
cross section of $pp\to W^\pm H^\mp+X$ at the LHC adopting the SUGRA-inspired
MSSM.
Sec.~\ref{sec:four} contains our conclusions.

\section{\label{sec:two}Analytic Results}

In this section, we express the $gg\to W^-H^+$ helicity amplitudes involving
one closed squark loop in terms of the standard scalar two-, three-, and
four-point functions,
\begin{eqnarray}
\lefteqn{B_0\left(p_1^2,m_0^2,m_1^2\right)
=\int\frac{d^Dq}{i\pi^2}\,\frac{1}{\left(q^2-m_0^2+i\epsilon\right)
\left[(q+p_1)^2-m_1^2+i\epsilon\right]},}
\nonumber\\
\lefteqn{C_0\left(p_1^2,(p_2-p_1)^2,p_2^2,m_0^2,m_1^2,m_2^2\right)}
\nonumber\\
&=&\int\frac{d^Dq}{i\pi^2}\,\frac{1}{\left(q^2-m_0^2+i\epsilon\right)
\left[(q+p_1)^2-m_1^2+i\epsilon\right]\left[(q+p_2)^2-m_2^2+i\epsilon\right]},
\nonumber\\
\lefteqn{D_0\left(p_1^2,(p_2-p_1)^2,(p_3-p_2)^2,p_3^2,p_2^2,(p_3-p_1)^2,
m_0^2,m_1^2,m_2^2,m_3^2\right)}
\nonumber\\
&=&\int\frac{d^Dq}{i\pi^2}\,\frac{1}{\left(q^2-m_0^2+i\epsilon\right)
\left[(q+p_1)^2-m_1^2+i\epsilon\right]\left[(q+p_2)^2-m_2^2+i\epsilon\right]
\left[(q+p_3)^2-m_3^2+i\epsilon\right]},
\nonumber\\
\end{eqnarray}
where $D$ is the space-time dimensionality.
The $B_0$ function is ultraviolet (UV) divergent in the physical limit
$D\to4$, while the $C_0$ and $D_0$ functions are UV finite in this limit.
We evaluate the $B_0$, $C_0$, and $D_0$ functions numerically with the aid of
the program package FF \cite{old}.
To simplify to notation, we introduce the abbreviations
$C_{ijk}^{ab}(c)=C_0\left(a,b,c,m_i^2,m_j^2,m_k^2\right)$ and
$D_{ijkl}^{abcd}(e,f)=D_0\left(a,b,c,d,e,f,m_i^2,m_j^2,m_k^2,m_l^2\right)$.

We work in the MSSM adopting the Feynman rules from Ref.~\cite{hab}.
For each quark flavor $q$ there is a corresponding squark flavor $\tilde q$,
which comes in two mass eigenstates $i=1,2$.
In the following, up- and down-type squark flavors are generically denoted by
$\tilde t$ and $\tilde b$, respectively.
The masses $m_{\tilde q_i}$ of the squarks and their trilinear couplings
$g_{W^-\tilde t_i\tilde b_j}$, $g_{h^0\tilde q_i\tilde q_j}$,
$g_{H^0\tilde q_i\tilde q_j}$, and $g_{H^+\tilde t_i\tilde b_j}$ to the 
$W^-$, $h^0$, $H^0$, and $H^+$ bosons are defined in Eqs.~(A.5) and (A.9) of
Ref.~\cite{hem} and in Eq.~(A.2) of Ref.~\cite{hh}, respectively.\footnote{%
In Ref.~\cite{hem}, $m_{\tilde q_i}$ and $g_{W^-\tilde t_i\tilde b_j}$ are
called $M_{\tilde Qa}$ and $\tilde V_{UaDb}^W/g$, respectively.}
Furthermore, we have
\begin{eqnarray}
g_{W^-H^+h^0}&=&-\frac{\cos(\alpha-\beta)}{2},
\nonumber\\
g_{W^-H^+H^0}&=&-\frac{\sin(\alpha-\beta)}{2},
\end{eqnarray}
where $\alpha$ is the mixing angle that rotates the weak CP-even Higgs
eigenstates into the mass eigenstates $h^0$ and $H^0$.

Calling the four-momenta of the two gluons and the $W$ boson $p_a$, $p_b$, and
$p_W$, respectively, we define the partonic Mandelstam variables as
$s=(p_a+p_b)^2$, $t=(p_a-p_W)^2$, and $u=(p_b-p_W)^2$.
Furthermore, we introduce the following short-hand notations: $w=m_W^2$,
$h=m_H^2$, $d=t-u$, $t_1=t-w$, $t_2=t-h$, $u_1=u-w$, $u_2=u-h$, $N=tu-wh$,
$\lambda=s^2+w^2+h^2-2(sw+wh+hs)$, and $q=m_{\tilde t_i}^2-m_{\tilde b_j}^2$.
We label the helicity states of the two gluons and the $W$ boson in the
partonic center-of-mass (c.m.) frame by $\lambda_a=-1/2,1/2$,
$\lambda_b=-1/2,1/2$, and $\lambda_W=-1,0,1$.

The relevant Feynman diagrams are depicted in Fig.~\ref{fig:one}.
In analogy to the quark case, we refer to the diagrams involving a neutral 
Higgs boson in the $s$ channel as triangle diagrams.
In contrast to the quark case, the diagrams involving the $A^0$ boson add up
to zero.
The residual diagrams are regarded to be of the box type.
The helicity amplitudes of the squark triangle contribution read
\begin{eqnarray}
\tilde{\cal M}^\triangle_{\lambda_a\lambda_b0}
&=&4\sqrt\lambda(1+\lambda_a\lambda_b)\sum_{\tilde q}\sum_i
\left(\frac{g_{W^-H^+h^0}g_{h^0\tilde q_i\tilde q_i}}
{s-m_{h^0}^2+im_{h^0}\Gamma_{h^0}}
+\frac{g_{W^-H^+H^0}g_{H^0\tilde q_i\tilde q_i}}
{s-m_{H^0}^2+im_{H^0}\Gamma_{H^0}}\right)
\nonumber\\
&&{}\times\left[1
+2m_{\tilde q_i}^2C^{00}_{\tilde q_i\tilde q_i\tilde q_i}(s)\right],
\nonumber\\
\tilde {\cal M}^\triangle_{\lambda_a\lambda_b\pm1}&=&0,
\end{eqnarray}
where $\Gamma_{h^0}$ and $\Gamma_{H^0}$ are the total decay widths of the 
$h^0$ and $H^0$ bosons, respectively.
In this case, the $W$ boson can only be longitudinally polarized because it
couples to two Higgs bosons.
As for the squark box contribution, all twelve helicity amplitudes contribute.
Due to Bose\footnote{%
Notice that the interchange of $t$ and $u$ also affects the representation of
the $W$-boson polarization four-vector through its dependence on the angle 
between the three-momenta of gluon $a$ and the $W$ boson.
This explains the sign factor in the first line of Eq.~(\ref{eq:m}), which is
not expected from pure Bose symmetry.}
and weak-isospin symmetry, they are related by
\begin{eqnarray}
\tilde{\cal M}_{\lambda_a\lambda_b\lambda_W}^\Box
\left(t,u,m_{\tilde t_i}^2,m_{\tilde b_j}^2\right)
&=&(-1)^{\lambda_W}\tilde{\cal M}_{\lambda_b\lambda_a\lambda_W}^\Box
\left(u,t,m_{\tilde t_i}^2,m_{\tilde b_j}^2\right),
\nonumber\\
\tilde{\cal M}_{\lambda_a\lambda_b\lambda_W}^\Box
\left(t,u,m_{\tilde t_i}^2,m_{\tilde b_j}^2\right)
&=&-\tilde{\cal M}_{-\lambda_a-\lambda_b-\lambda_W}^\Box
\left(t,u,m_{\tilde b_j}^2,m_{\tilde t_i}^2\right),
\label{eq:m}
\end{eqnarray}
respectively.
Keeping $\lambda_W=\pm1$ generic, we thus only need to specify four
expressions.
These read:
\begin{eqnarray}
\tilde{\cal M}_{++0}^\Box&=&\frac{4}{s\sqrt\lambda}
\sum_{\left(\tilde t,\tilde b\right)}\sum_{i,j}
g_{W^-\tilde t_i\tilde b_j}g_{H^+\tilde t_i\tilde b_j}
\left[\tilde F_{++}^0+(t\leftrightarrow u)\right],
\nonumber\\
\tilde{\cal M}_{+-0}^\Box&=&\frac{4}{N\sqrt\lambda}
\sum_{\left(\tilde t,\tilde b\right)}\sum_{i,j}
g_{W^-\tilde t_i\tilde b_j}g_{H^+\tilde t_i\tilde b_j}
\left[\tilde F_{+-}^0
-\left(m^2_{\tilde t_i}\leftrightarrow m^2_{\tilde b_j}\right)\right],
\nonumber\\
\tilde{\cal M}_{++\lambda_w}^\Box
&=&\frac{m_W}{\sqrt{N}}\left(\frac{2}{s}\right)^{3/2}
\sum_{\left(\tilde t,\tilde b\right)}\sum_{i,j}
g_{W^-\tilde t_i\tilde b_j}g_{H^+\tilde t_i\tilde b_j}
\left[\left(\frac{\tilde F_{++}^1}{\sqrt\lambda}
+\lambda_W\tilde F_{++}^2\right)-(t\leftrightarrow u)\right],
\nonumber\\
\tilde{\cal M}_{+-\lambda_W}^\Box
&=&\frac{m_W}{\sqrt{s}}\left(\frac{2}{N}\right)^{3/2}
\sum_{\left(\tilde t,\tilde b\right)}\sum_{i,j}
g_{W^-\tilde t_i\tilde b_j}g_{H^+\tilde t_i\tilde b_j}
\left[\left(\frac{\tilde F_{+-}^1}{\sqrt\lambda}
+\lambda_W\tilde F_{+-}^2\right)
-\left(m^2_{\tilde t_i}\leftrightarrow m^2_{\tilde b_j}\right)\right],
\label{eq:b}
\end{eqnarray}
where $\sum_{\left(\tilde t,\tilde b\right)}$ denotes the sum over squark
generations and
\begin{eqnarray}
\tilde F_{++}^0&=& 
2s(t_1+u_1)\left[m_{\tilde b_j}^2C^{00}_{\tilde b_j\tilde b_j\tilde b_j}(s)
-m_{\tilde t_i}^2C^{00}_{\tilde t_i \tilde t_i\tilde t_i}(s)\right]
\nonumber\\
&&{}+[wd-q(t_1+u_1)]\left[t_2C^{h0}_{\tilde b_j\tilde t_i\tilde t_i}(t)
+t_1C^{w0}_{\tilde t_i\tilde b_j\tilde b_j}(t)\right]
\nonumber\\
&&{}-[wd+q(t_1+u_1)]\left[t_2C^{h0}_{\tilde t_i\tilde b_j\tilde b_j}(t)
+t_1C^{w0}_{\tilde b_j\tilde t_i\tilde t_i}(t)\right]
\nonumber\\
&&{}-[wd-q(t_1+u_1)]
\left[N+s\left(m_{\tilde b_j}^2+m_{\tilde t_i}^2\right)\right]
D^{h0w0}_{\tilde b_j\tilde t_i\tilde t_i\tilde b_j}(t,u) 
\nonumber\\
&&{}+2sm_{\tilde b_j}^2[w(t_2+u_2)+q(t_1+u_1)]
D^{hw00}_{\tilde b_j\tilde t_i\tilde b_j\tilde b_j}(s,t)
\nonumber\\
&&{}-2sm_{\tilde t_i}^2[w(t_2+u_2)-q(t_1+u_1)]
D^{hw00}_{\tilde t_i\tilde b_j\tilde t_i\tilde t_i}(s,t),
\nonumber \\
\tilde F_{+-}^0&=&
s(t+u-2q)[w(t_2+u_2)+q(t_1+u_1)]C^{00}_{\tilde b_j\tilde b_j\tilde b_j}(s)
\nonumber\\
&&{}-t_2\{tw(t_2+u_2)-q[t(t+3u)-2w(t+u)-2N]\}
C^{h0}_{\tilde b_j\tilde t_i\tilde t_i}(t) 
\nonumber\\
&&{}-u_2\{uw(t_2+u_2)-q[u(3t+u)-2w(t+u)-2N]\}
C^{h0}_{\tilde b_j\tilde t_i\tilde t_i}(u)
\nonumber\\
&&{}-t_1\{tw(t_2+u_2)-q[t(d+4u_1)-2N]\}
C^{w0}_{\tilde b_j\tilde t_i\tilde t_i}(t)
\nonumber\\
&&{}-u_1\{uw(t_2+u_2)+q[u(d-4t_1)+2N]\}
C^{w0}_{\tilde b_j\tilde t_i\tilde t_i}(u)
\nonumber\\
&&{}-(d^2+2N)[w(t_2+u_2)+q(t_1+u_1)]C^{hw}_{\tilde b_j\tilde t_i\tilde b_j}(s)
\nonumber\\
&&{}+[wd-q(t_1+u_1)]
\left[N\left(m_{\tilde b_j}^2+m_{\tilde t_i}^2\right)+sq^2\right]
D^{h0w0}_{\tilde b_j \tilde t_i\tilde t_i\tilde b_j}(t,u)
\nonumber\\
&&{}-[w(t_2+u_2)+q(t_1 + u_1)]
\left[st\left(t-2m_{\tilde t_i}^2\right)-2t_1t_2m_{\tilde b_j}^2+sq^2\right]
D^{hw00}_{\tilde b_j\tilde t_i\tilde b_j\tilde b_j}(s,t)
\nonumber\\
&&{}-[w(t_2+u_2)+q(t_1+u_1)]
\left[su\left(u-2m_{\tilde t_i}^2\right)-2u_1u_2m_{\tilde b_j}^2+sq^2\right]
D^{hw00}_{\tilde b_j\tilde t_i\tilde b_j\tilde b_j}(s,u),
\nonumber\\
\tilde F_{++}^1&=&
-2s^2d\left[m_{\tilde b_j}^2C^{00}_{\tilde b_j\tilde b_j\tilde b_j}(s)
-m_{\tilde t_i}^2C^{00}_{\tilde t_i \tilde t_i\tilde t_i}(s)\right]
\nonumber\\
&&{}+[N(t_1+u_1)+sdq]\left[t_2C^{h0}_{\tilde b_j\tilde t_i\tilde t_i}(t)
+t_1C^{w0}_{\tilde t_i\tilde b_j\tilde b_j}(t)\right]
\nonumber\\
&&{}-[N(t_1+u_1)-sdq]\left[t_2C^{h0}_{\tilde t_i\tilde b_j\tilde b_j}(t)
+t_1C^{w0}_{\tilde b_j\tilde t_i\tilde t_i}(t)\right]
\nonumber\\
&&{}-\left[N+s(m_{\tilde b_j}^2+m_{\tilde t_i}^2)\right]
\left[N(t_1+u_1)+sdq\right]
D^{h0w0}_{\tilde b_j\tilde t_i\tilde t_i\tilde b_j}(t,u)
\nonumber\\
&&{}+2s^2m_{\tilde b_j}^2[2N+d(t-q)]
D^{hw00}_{\tilde b_j\tilde t_i\tilde b_j\tilde b_j}(s,t)
-2s^2m_{\tilde t_i}^2[2N+d(t+q)]
D^{hw00}_{\tilde t_i\tilde b_j\tilde t_i\tilde t_i}(s,t),
\nonumber\\
\tilde F_{++}^2&=&
(N-sq)\left[t_2C^{h0}_{\tilde b_j\tilde t_i\tilde t_i}(t)
-t_1C^{w0}_{\tilde b_j\tilde t_i\tilde t_i}(t)\right] 
-(N+sq)\left[t_2C^{h0}_{\tilde t_i\tilde b_j\tilde b_j}(t)
-t_1C^{w0}_{\tilde t_i\tilde b_j\tilde b_j}(t)\right]
\nonumber\\ 
&&{}-\left[N^2+2sN\left(m_{\tilde b_j}^2+m_{\tilde t_i}^2\right)+s^2q^2\right]
D^{h0w0}_{\tilde b_j \tilde t_i\tilde t_i\tilde b_j}(t,u),
\nonumber\\
\tilde F_{+-}^1&=&
2sdNB_0\left(s,m_{\tilde b_j}^2,m_{\tilde b_j}^2\right)
+s^2d\left[t^2+u^2+2N-2q(t+u)+2q^2\right]
C^{00}_{\tilde b_j\tilde b_j\tilde b_j}(s)
\nonumber\\
&&{}-t_2\{st(td+2N)+q[d(2st+3N)-2N(2t_1+t_2)]\}
C^{h0}_{\tilde b_j\tilde t_i\tilde t_i}(t) 
\nonumber\\
&&{}-u_2\{su(ud-2N)+q[d(2su+3N)+2N(2u_1+u_2)]\}
C^{h0}_{\tilde b_j\tilde t_i\tilde t_i}(u)
\nonumber\\
&&{}-t_1\{st(td+2N)+q[d(2st+N)-2t_2N]\}
C^{w0}_{\tilde b_j\tilde t_i\tilde t_i}(t)
\nonumber\\
&&{}-u_1\{su(ud-2N)+q[d(2su+N)+2u_2N]\}
C^{w0}_{\tilde b_j\tilde t_i\tilde t_i}(u)
\nonumber\\
&&{}-sd(d^2+4N)(t+u-2q)
C^{hw}_{\tilde b_j\tilde t_i\tilde b_j}(s)
\nonumber\\
&&{}+[2t_1N-d(N-sq)]
\left[N\left(m_{\tilde b_j}^2+m_{\tilde t_i}^2\right)+sq^2\right]
D^{h0w0}_{\tilde b_j \tilde t_i\tilde t_i\tilde b_j}(t,u)
\nonumber\\
&&{}-s[2N+d(t-q)]\left[st^2+2Nm_{\tilde b_j}^2-sq(2t-q)\right]
D^{hw00}_{\tilde b_j\tilde t_i\tilde b_j\tilde b_j}(s,t)
\nonumber\\
&&{}+s[2N-d(u-q)]\left[su^2+2Nm_{\tilde b_j}^2-sq(2u-q)\right]
D^{hw00}_{\tilde b_j\tilde t_i\tilde b_j\tilde b_j}(s,u),
\nonumber \\
\tilde F_{+-}^2&=&
2sNB_0\left(s,m_{\tilde b_j}^2,m_{\tilde b_j}^2\right) 
+s\left[s(t^2+u^2)+4Nm_{\tilde b_j}^2-2sq(t+u-q)\right]
C^{00}_{\tilde b_j\tilde b_j\tilde b_j}(s)
\nonumber\\
&&{}-t_2[st^2+q(2st+N)]C^{h0}_{\tilde b_j\tilde t_i\tilde t_i}(t)
-u_2[su^2+q(2su+N)]C^{h0}_{\tilde b_j\tilde t_i\tilde t_i}(u)
\nonumber\\
&&{}-t_1[st^2+q(2st+N)]C^{w0}_{\tilde b_j\tilde t_i\tilde t_i}(t)
-u_1[su^2+q(2su+N)]C^{w0}_{\tilde b_j\tilde t_i\tilde t_i}(u)
\nonumber\\
&&{}-s(d^2+2N)(t+u-2q)C^{hw}_{\tilde b_j\tilde t_i\tilde b_j}(s)
\nonumber\\
&&{}+q\left\{N\left[N+2s(m_{\tilde b_j}^2+m_{\tilde t_i}^2)\right]+s^2q^2
\right\}D^{h0w0}_{\tilde b_j \tilde t_i\tilde t_i\tilde b_j}(t,u)
\nonumber\\
&&{}-s(t-q)\left[st^2+4Nm_{\tilde b_j}^2-sq(2t-q)\right]
D^{hw00}_{\tilde b_j\tilde t_i\tilde b_j\tilde b_j}(s,t)
\nonumber\\
&&{}-s(u-q)\left[su^2+4Nm_{\tilde b_j}^2-sq(2u-q)\right]
D^{hw00}_{\tilde b_j\tilde t_i\tilde b_j\tilde b_j}(s,u).
\end{eqnarray}
Notice that the UV divergences of $\tilde F_{+-}^1$ and $\tilde F_{+-}^2$
cancel in the expression for $\tilde{\cal M}_{+-\lambda_W}^\Box$ in
Eq.~(\ref{eq:b}).

The differential cross section of the partonic subprocess $gg\to W^-H^+$ is
then given by \cite{box}
\begin{eqnarray}
\frac{d\sigma}{dt}(gg\to W^-H^+)&=&\frac{\alpha_s^2(\mu_r)G_F^2m_W^2}
{256(4\pi)^3s^2}\sum_{\lambda_a,\lambda_b,\lambda_W}\left|
{\cal M}_{\lambda_a\lambda_b\lambda_W}^\triangle+
{\cal M}_{\lambda_a\lambda_b\lambda_W}^\Box\right.
\nonumber\\
&&{}-\left.\tilde{\cal M}_{\lambda_a\lambda_b\lambda_W}^\triangle
-\tilde{\cal M}_{\lambda_a\lambda_b\lambda_W}^\Box\right|^2,
\label{eq:x}
\end{eqnarray}
where $\alpha_s(\mu_r)$ is the strong-coupling constant at renormalization
scale $\mu_r$, $G_F$ is Fermi's constant, and
${\cal M}_{\lambda_a\lambda_b\lambda_W}^\triangle$ and
${\cal M}_{\lambda_a\lambda_b\lambda_W}^\Box$ are the helicity amplitudes of
the quark triangle and box contributions, which may be found in Eqs.~(1) and
(3) of Ref.~\cite{box}, respectively.
The relative minus signs between the quark and squark terms in
Eq.~(\ref{eq:x}) compensate for the fact that the Feynman rules underlying
Ref.~\cite{box} differ from those adopted here.
Due to Bose symmetry, the cross section $d\sigma/dt$ of $gg\to W^-H^+$ is
symmetric in $t$ and $u$.
Due to charge-conjugation invariance, it coincides with the one of
$gg\to W^+H^-$, so that the cross section $d\sigma/dt$ of $gg\to W^\pm H^\mp$
emerges from the right-hand side of Eq.~(\ref{eq:x}) by multiplication with 
two.
The kinematics of the inclusive reaction $AB\to W^\pm H^\mp+X$, where $A$ and
$B$ are colliding hadrons, is described in Sec.~II of Ref.~\cite{wh}.
Its double-differential cross section $d^2\sigma/dy\,dp_T$, where $y$ and
$p_T$ are the rapidity and transverse momentum of the $W$ boson in the c.m.\
system of the hadronic collision, may be evaluated from Eq.~(2.1) of
Ref.~\cite{wh}.

\section{\label{sec:three}Phenomenological Implications}

We are now in a position to explore the phenomenological implications of our
results.
The SM input parameters for our numerical analysis are taken to be
$G_F=1.16639\cdot10^{-5}$~GeV$^{-2}$, $m_W=80.419$~GeV, $m_Z=91.1882$~GeV,
$m_t=174.3$~GeV , and $m_b=4.6$~GeV \cite{pdg}.
We adopt the lowest-order set CTEQ5L \cite{lai} of parton density functions 
for the proton.
We evaluate $\alpha_s(\mu_r)$ from the lowest-order formula \cite{pdg} with
$n_f=5$ quark flavors and asymptotic scale parameter
$\Lambda_{\rm QCD}^{(5)}=146$~MeV \cite{lai}.
We identify the renormalization and factorization scales with the
$W^\pm H^\mp$ invariant mass $s$.
For our purposes, it is useful to replace $m_A$ by $m_H$, the mass of the
$H^\pm$ bosons to be produced, in the set of MSSM input parameters.
We vary $\tan\beta$ and $m_H$ in the ranges $2.5<\tan\beta<38\approx m_t/m_b$
and 120~GeV${}<m_H<600$~GeV, respectively.
As for the GUT parameters, we choose $m_{1/2}=150$~GeV, $A=0$, and $\mu<0$, 
and tune $m_0$ so as to be consistent with the desired value of $m_H$.
All other MSSM parameters are then determined according to the SUGRA-inspired
scenario as implemented in the program package SUSPECT \cite{djo}.
We do not impose the unification of the $\tau$-lepton and $b$-quark Yukawa
couplings at the GUT scale, which would just constrain the allowed $\tan\beta$
range without any visible effect on the results for these values of
$\tan\beta$.
We exclude solutions which do not comply with the present experimental lower
mass bounds of the sfermions, charginos, neutralinos, and Higgs bosons
\cite{ruh}.

\begin{figure}[ht]
\begin{center}
\begin{tabular}{cc}
\parbox{8cm}{\epsfig{figure=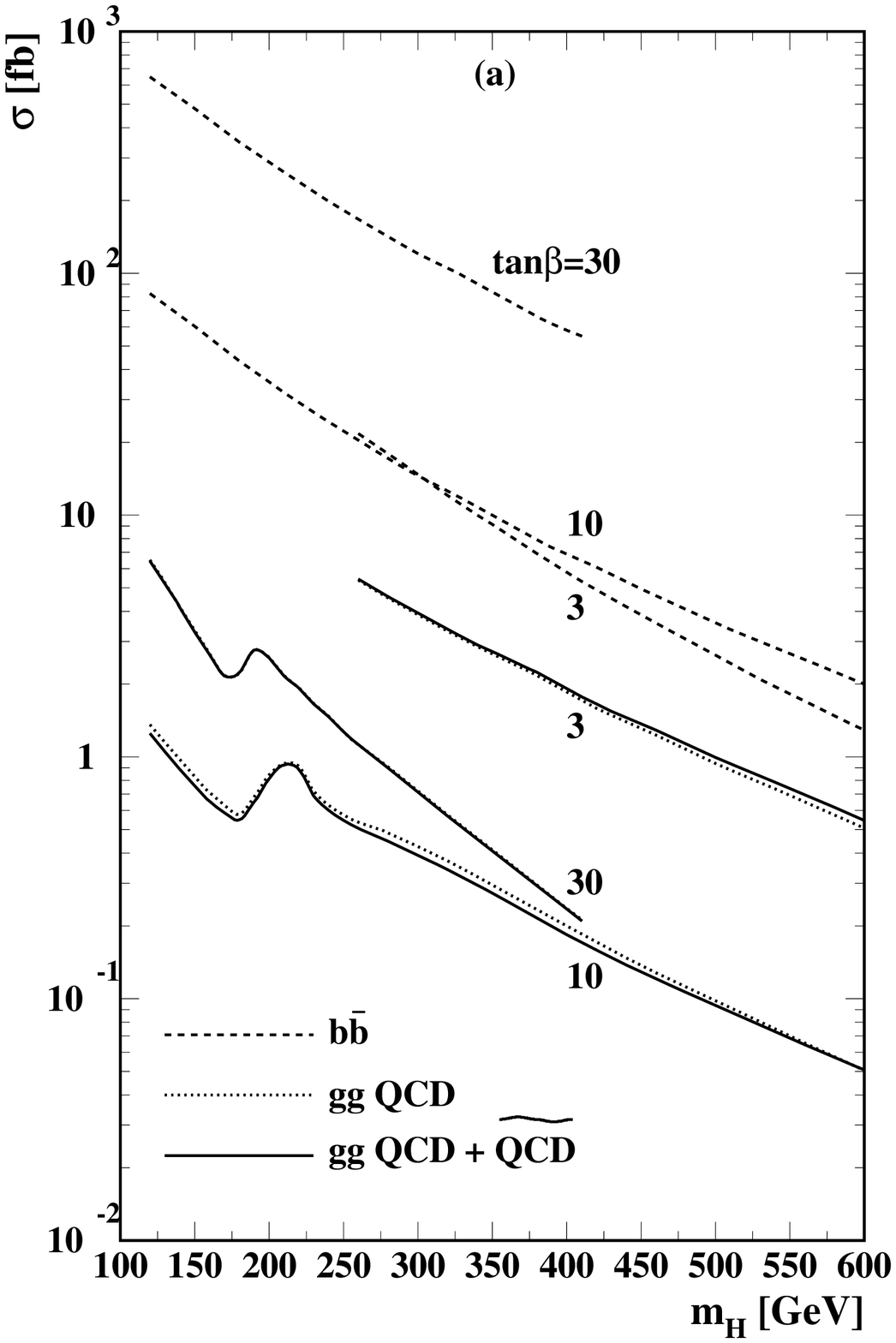,width=8cm}}
&
\parbox{8cm}{\epsfig{figure=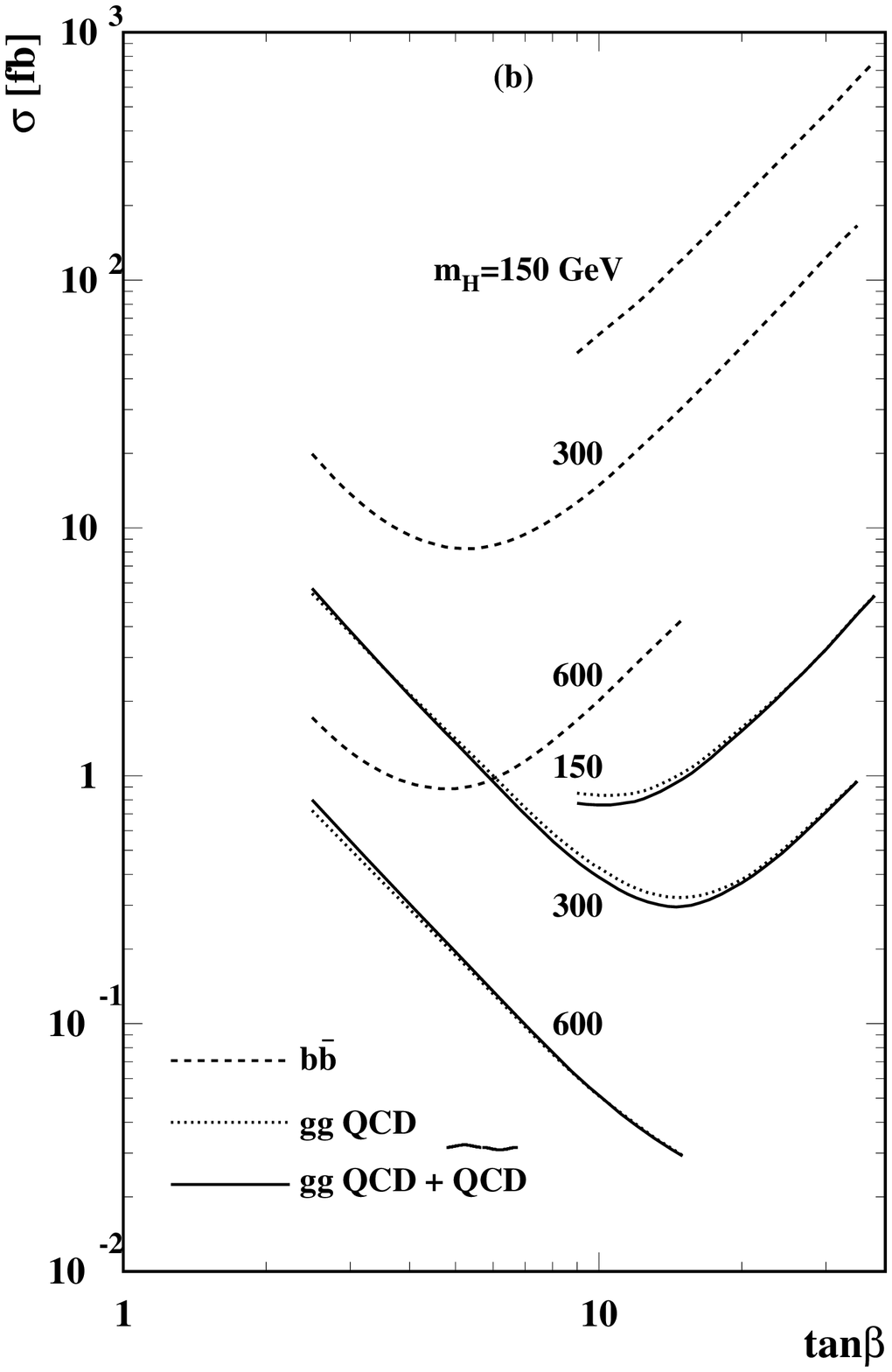,width=8cm}}
\end{tabular}
\caption{Total cross sections $\sigma$ (in fb) of $pp\to W^\pm H^\mp+X$ via
$b\bar b$ annihilation (dashed lines) and $gg$ fusion (solid lines) at the LHC
(a) as functions of $m_H$ for $\tan\beta=3$, 10, and 30; and
(b) as functions of $\tan\beta$ for $m_H=150$, 300, and 600~GeV.
For comparison, also the quark loop contribution to $gg$ fusion (dotted lines)
is shown.}
\label{fig:two}
\end{center}
\end{figure}

We now study $pp\to W^\pm H^\mp+X$ at the LHC, with c.m.\ energy
$\sqrt S=14$~TeV.
The fully integrated cross section is considered as a function of $m_H$ for
$\tan\beta=3$, 10, and 30 in Fig.~\ref{fig:two}(a) and as a function of
$\tan\beta$ for $m_H=150$, 300, and 600~GeV in Fig.~\ref{fig:two}(b).
The combined $gg$-fusion contribution due to quarks and squarks (solid lines)
is compared with the one due to quarks only (dotted lines) \cite{wh}.
For reference, also the $b\bar b$-annihilation contribution (dashed lines) is
shown \cite{wh}.
We note that the SUGRA-inspired MSSM with our choice of input parameters does
not permit $\tan\beta$ and $m_H$ to be simultaneously small, due to the
experimental selectron mass lower bound \cite{ruh}.
This explains why the curves for $\tan\beta=3$ in Fig.~\ref{fig:two}(a) only
start at $m_H\approx260$~GeV and those for $m_H=150$~GeV in
Fig.~\ref{fig:two}(b) at $\tan\beta\approx9$.
For large $m_H$, the experimental $m_h$ lower bound \cite{ruh} enforces
$\tan\beta\agt2.5$.
On the other hand, the experimental lower bounds on the chargino and
neutralino masses \cite{ruh} induce an upper limit on $\tan\beta$, which
depends on $m_H$.
We observe from Figs.~\ref{fig:two}(a) and (b) that the supersymmetric
correction to the $gg$-fusion cross section can be of either sign and have a
magnitude of order 10\%.
It exceeds $+10\%$ for small $\tan\beta$ and large $m_H$, while it almost 
reaches $-10\%$ for medium $\tan\beta$ and small or medium $m_H$.
On the other hand, it is generally small for large $\tan\beta$.
We recall that, in the case of $pp\to H^+H^-+X$, the supersymmetric correction
to the $gg$-fusion cross section can be as large as $+50\%$ \cite{hh}.
As explained in Ref.~\cite{wh}, the dip in the $m_H$ dependence of the 
$gg$-fusion cross section located about $m_H=m_t$ [see Fig.~\ref{fig:two}(a)]
arises from resonating top-quark propagators in the quark box diagrams.
Furthermore, the minima of the curves in Fig.~\ref{fig:two}(b) close to
$\tan\beta\approx\sqrt{m_t/m_b}\approx6$ may be understood by observing that
the average strength of the $H^-\bar bt$ coupling, which is proportional to
$\sqrt{m_t^2\cot^2\beta+m_b^2\tan^2\beta}$, is then minimal \cite{hh}.
As is the quark case \cite{wh}, the squark triangle and box contributions are
similar in size and destructively interfere with each other, so that their
superposition is much smaller than each of them separately.
As in the case of $pp\to H^+H^-+X$ \cite{hh}, the bulk of the squark
contribution comes from the stop and sbottom squarks, while the contributions
from the first- and second-generation squarks is greatly suppressed because
their couplings to the Higgs bosons are significantly smaller than those of
the third-generation squarks and their masses are generally larger than those
of lightest stop and sbottom squarks, $\tilde t_1$ and $\tilde b_1$.
We conclude that the suppression of the $gg$-fusion cross section relative to
the one of $b\bar b$ annihilation remains after the inclusion of the squark 
loop contributions.

\begin{figure}[ht]
\begin{center}
\begin{tabular}{cc}
\parbox{8cm}{\epsfig{figure=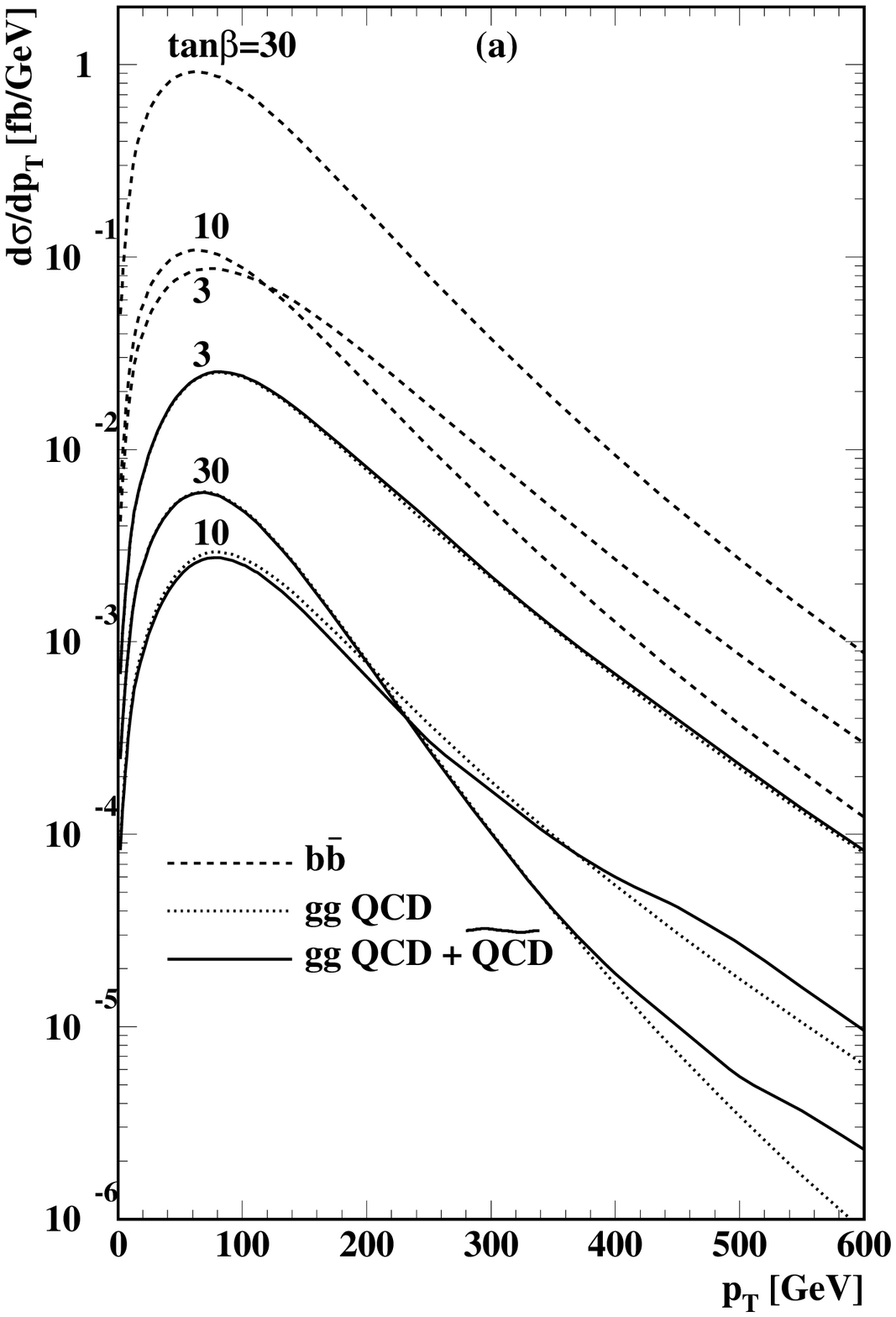,width=8cm}}
&
\parbox{8cm}{\epsfig{figure=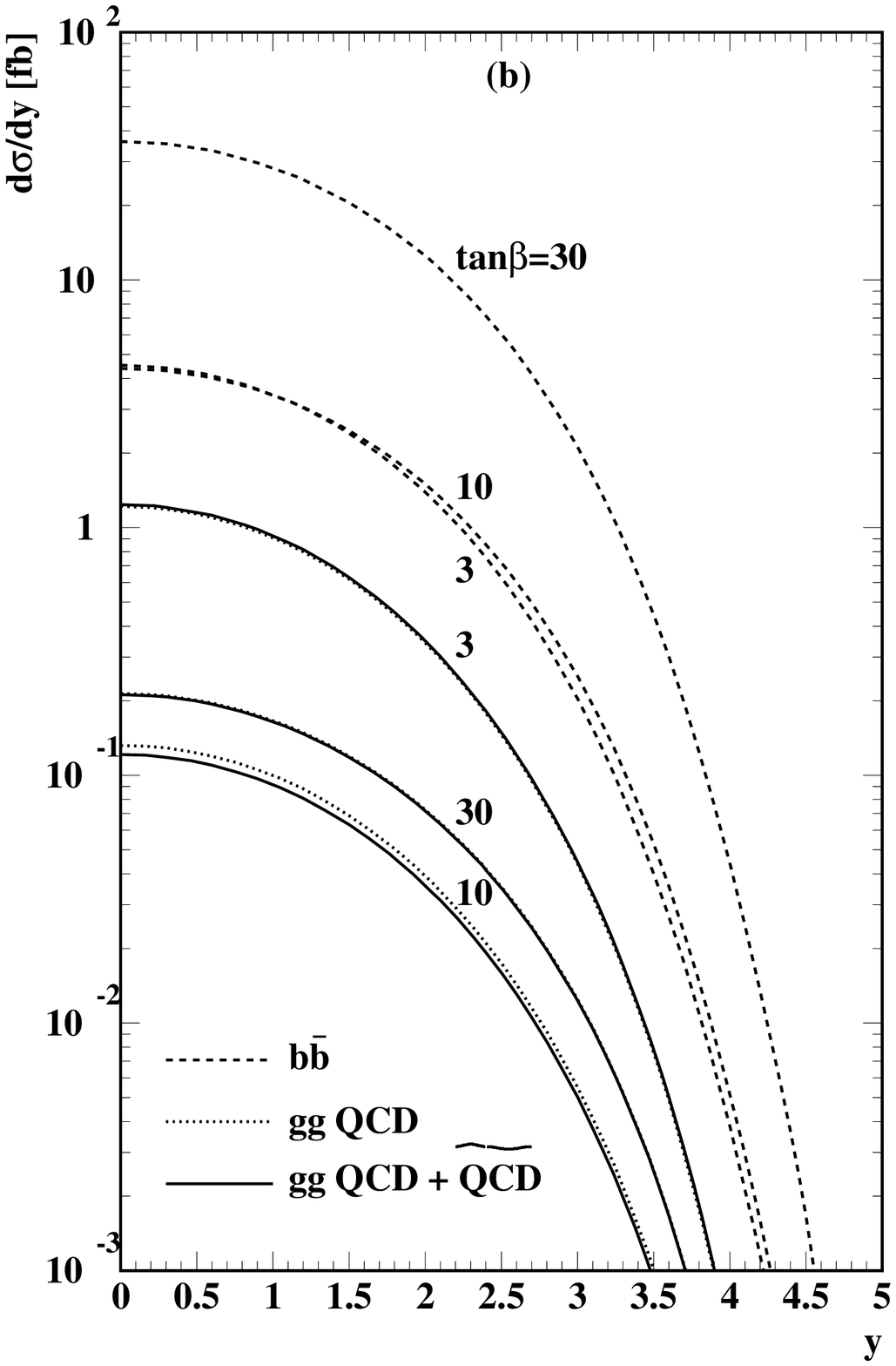,width=8cm}}
\end{tabular}
\caption{(a) $p_T$ distributions $d\sigma/dp_T$ (in fb/GeV) and (b) $y$
distributions $d\sigma/dy$ (in fb) of $pp\to W^\pm H^\mp+X$ via $b\bar b$
annihilation (dashed lines) and $gg$ fusion (solid lines) at the LHC for
$\tan\beta=3$, 10, 30, and $m_H=300$~GeV.
For comparison, also the quark loop contribution to $gg$ fusion (dotted lines)
is shown.}
\label{fig:three}
\end{center}
\end{figure}

It is interesting to find out how the kinematic behavior of the $gg$-fusion
cross section is affected by the supersymmetric correction.
To that end, we study in Figs.~\ref{fig:three}(a) and (b) the distributions in
the $W$-boson transverse momentum $p_T$ and rapidity $y$, respectively, for
$\tan\beta=3$, 10, 30, and $m_H=300$~GeV.
While the $y$ distribution does not exhibit any striking features, we observe 
that the supersymmetric correction leads to an increase of the $p_T$ 
distribution by more than 50\% at large $p_T$ for medium to large $\tan\beta$.
This can be traced to the presence of absorptive parts in the squark loop
contribution.
In fact, if
$p_T>\sqrt{\lambda\left(4m_{\tilde q_i}^2,m_W^2,m_H^2\right)}/
\left(4m_{\tilde q_i}\right)$, then $s>2m_{\tilde q_i}$, so that pairs of
real $\tilde q_i$ squarks can be produced.

For a comparison with future experimental data, the $b\bar b$-annihilation and
$gg$-fusion channels should be combined.
From Fig.~\ref{fig:two}(a), we read off that the total cross section of
$pp\to W^\pm H^\mp+X$ at the LHC is predicted to be approximately 500~fb
(20~fb) in the considered MSSM scenario if $\tan\beta=30$ and $m_H=150$~GeV
($\tan\beta=3$ and $m_H=300$~GeV).
If we assume the integrated luminosity per year to be at its design value of
$L=100$~fb$^{-1}$ for each of the two LHC experiments, ATLAS and CMS, then
this translates into about 100.000 (4.000) signal events per year.

\section{\label{sec:four}Conclusions}

We calculated the squark loop contribution to the partonic subprocess
$gg\to W^\pm H^\mp$ within the MSSM, and analyzed its impact on the inclusive
cross section of $pp\to W^\pm H^\mp+X$ and its distributions in transverse
momentum and rapidity at the LHC adopting a SUGRA-inspired scenario.
Its inclusion may increase or decrease the integrated $gg$-fusion cross
section by up to 10\%, depending on the values $\tan\beta$ and $m_H$.
However, $b\bar b$ annihilation remains the dominant mechanism of
$W^\pm H^\mp$ associated hadroproduction at the LHC.
Should the MSSM be realized in nature, then the $W^\pm H^\mp$ channel will 
provide a copious source of charged Higgs bosons at the LHC, with an annual
yield of up to 100.000 signal events.

\vspace{1cm}
\noindent
{\bf Acknowledgements}
\smallskip

We thank Stefan Dittmaier for useful communications regarding the sign factor
in Eq.~(\ref{eq:m}).
B.A.K. thanks the Theory Group of the Werner-Heisenberg-Institut for the
hospitality extended to him during a visit when this paper was finalized.
The work of A.A.B.B. was supported by the Friedrich-Ebert-Stiftung through
Grant No.~219747.
The work of B.A.K. was supported in part by the Deutsche
Forschungsgemeinschaft through Grant No.\ KN~365/1-1, by the
Bundesministerium f\"ur Bildung und Forschung through Grant No.\ 05~HT9GUA~3,
and by the European Commission through the Research Training Network
{\it Quantum Chromodynamics and the Deep Structure of Elementary Particles}
under Contract No.\ ERBFMRX-CT98-0194.

\newpage

\newpage

\newpage

\end{document}